\documentclass[twocolumn,showpacs,preprintnumbers,amsmath,amssymb,prb]{revtex4}
\usepackage{graphicx} \usepackage{dcolumn} \usepackage{bm}
\usepackage{subfigure} \usepackage{float} \usepackage{color}

\begin{document}

\title{Thermodynamics of the two-dimensional frustrated $J_1$-$J_2$ Heisenberg ferromagnet in
the collinear stripe regime: Susceptibility and correlation length }

\author{M. H\"{a}rtel}
\author{J. Richter}
\author{O. G\"otze}
\affiliation{Institut f\"{u}r Theoretische Physik, Otto-von-Guericke-Universit\"{a}t Magdeburg, D-39016 Magdeburg, Germany}
\author{D. Ihle}
\affiliation{Institut f\"{u}r Theoretische Physik, Universit\"{a}t Leipzig, D-04109 Leipzig, Germany}
\author{S.-L. Drechsler}
\affiliation{Leibniz-Institut f\"{u}r Festk\"{o}rper- und Werkstoffforschung
Dresden, D-01171 Dresden, Germany}

\date{\today}

\begin{abstract}
We calculate the temperature dependence of the correlation
length $\xi$ and the uniform susceptibility $\chi_0$ of the frustrated
$J_1$-$J_2$ square-lattice Heisenberg ferromagnet in
the collinear stripe phase using Green-function technique. 
The height $\chi_{max}$ and  the 
position $T(\chi_{max})$ of the maximum in the $\chi_0(T)$ curve exhibit a
characteristic 
dependence on the frustration parameter $J_2/|J_1|$, which is well
described, for $J_2 > 0.7|J_1|$, 
by the relations  $\chi_{max}=a \left(J_2-J_2^c\right)^{-\nu}$ and 
$T(\chi_{max})=b(J_2-J_2^c)$, where $J_2^c = 0.4|J_1|$ and  $\nu$ is of the
order of unity.
The correlation length diverges at low temperatures as $\xi \propto
e^{A/T}$, where $A$ increases  with growing $J_2/|J_1|$.  
    We also compare our results with recent measurements on 
layered oxovanadates and find reasonable agreement.
\end{abstract}
\maketitle
\section{\label{sec:level1}Introduction}

Frustrated square-lattice quantum magnets 
have been in the focus of active condensed-matter investigations in recent
years.
While for the study of ground state (GS) properties many alternative methods, such as
exact diagonalization (ED),\cite{schulz,shannon06,ED40,richter10} the  Schwinger
boson approach,\cite{feldner11}
functional renormalization group method,\cite{reuther10} the tensor-product 
approach,\cite{Verstraete:2011} or the coupled-cluster
method (CCM)\cite{richter10}
can be applied, there are much less reasonable theoretical approaches available to deal with
thermodynamic properties of these systems.
On the other hand, there are many recent experimental studies on quasi-two-dimensional
frustrated
square-lattice compounds, see, e.g.,
Refs.~\onlinecite{kaul04,enderle,nath,rosner09,rosner09a,nath08,carretta2009,carretta2011,tsirlin2011,johnston2011},
where typically temperature dependent properties are reported, which should
be compared with theoretical predictions.
The quantum Monte Carlo technique is not applicable due to the sign
problem for frustrated systems.\cite{troyer2005} 
The high-temperature expansion approach\cite{Rosner_HTE,schmidt11}
is limited to temperatures down to the order of the exchange coupling.  
Full ED studies used in 
Refs.~\onlinecite{shannon04,schmidt07,schmidt07_2} 
for the $J_1$-$J_2$ square-lattice Heisenberg
ferromagnet of $N=8,16$ and $20$ sites
suffer from finite-size effects at lower
temperatures.\cite{haertel10,schmidt11}  
An alternative method to describe quantum magnets
in the whole temperature
range is the  Green-function technique.\cite{elk,froebrich06,rudoy2011} A rotationally invariant
second-order Green-function theory has been applied successfully to
describe the thermodynamics of frustrated quantum
magnets.\cite{barabanov94,ihle2001,barabanov2009,haertel08,haertel10,barabanov2011} 
In particular, the Green-function technique is designed for $N
\to \infty$ and allows the
 calculation of  the magnetic correlation length, in addition to the usual thermodynamic
 quantities, such as the susceptibility.

Motivated by recent measurements of the correlation lengths for several frustrated
layered Heisenberg
square-lattice ferromagnets,\cite{carretta2011}
in the present paper we study the temperature dependence of the correlation length and the
uniform susceptibility
of the spin-1/2 $J_1$-$J_2$ model 
\begin{equation}
  H=J_1\sum_{\langle i,j\rangle}{\bf S}_i{\bf S}_j
+J_2\sum_{\langle\langle i,j\rangle\rangle}{\bf S}_i{\bf S}_j ,
\end{equation}
where $\langle \ldots\rangle$ denotes the nearest neighbor (NN) and $\langle\langle \ldots\rangle\rangle$ 
the next-nearest neighbor (NNN) bonds
on a square lattice. We consider a ferromagnetic NN coupling  $J_1<0$ and
a frustrating antiferromagnetic NNN coupling $J_2 > 0$.
The GS of this model has been discussed in
Refs.~\onlinecite{shannon06,richter10,feldner11}.  At $J_2 = J_2^c \approx 0.4 |J_1|$
the ferromagnetic GS
present at small $J_2$ gives way for a GS phase with zero
total magnetization. Although, the nature of this state for  $J_2 \gtrsim J_2^c$
is still under debate, the existence of antiferromagnetic collinear stripe
GS long-range order (LRO) for $J_2 > 0.7 |J_1|$ is not questioned.

Following our previous investigation of this model, see
Ref.~\onlinecite{haertel10}, we use the rotationally invariant
second-order Green-function method (RGM) to calculate the thermodynamic
properties.
However, by contrast to Ref.~\onlinecite{haertel10}, where the model was
studied in the ferromagnetic regime, i.e.
$J_2 < J_2^c$,  here we focus on the antiferromagnetic collinear stripe
GS regime. We restrict our study on the parameter region $J_2 >
0.7 |J_1|$, that
corresponds to the experimental situation  for several layered
oxovanadates.\cite{kaul04,nath08,carretta2011}

The paper is organized as follows. 
In Sec.~\ref{sec2} the RGM applied to the $J_1$-$J_2$ square lattice is
illustrated. 
The results for the susceptibility and correlation length and the comparison to recent 
experimental data is presented in Sec. \ref{secIII}. Finally, a short summary is given in Sec. \ref{secIV}.

\section{Rotationally invariant Green-function method (RGM)}\label{sec2}
The RGM was introduced by Kondo and Yamaji.\cite{kondo} 
The method was further developed and  applied to Heisenberg magnets by several 
groups, see, e.g.,
Refs.~\onlinecite{shimahara91,SSI94,barabanov94,winterfeldt97,schindelin1999,Yu2000,ihle2001,canals2002,schmal2004,schmal06,barabanov2009,haertel08,junger3d,ihle2008,haertel10,barabanov2011}.  \\

To calculate the dynamical transverse spin susceptibility
$\chi_q^{+-}\left(\omega\right)$ we have to determine the two-time commutator
Green function $\langle\langle S_q^+;S_{-q}^-\rangle\rangle_\omega=-\chi_q^{+-}\left(\omega\right)$.
The equation of motion for  $\langle\langle
S_q^+;S_{-q}^-\rangle\rangle_\omega$ up to second order reads 
$\omega^2 \langle\langle S_q^+;S_{-q}^-\rangle\rangle_\omega=
M_q+\langle\langle -\ddot{S}_q^+;S_{-q}^-\rangle\rangle_\omega$ with $-\ddot{S}_q^+=\left[\left[S_q^+,H\right],H\right]$ 
and  the exact moment
\begin{equation}
M_q=-8J_1C_{1,0}\left(1-\gamma_q^{(1)}\right)-8J_2C_{1,1}\left(1-\gamma_q^{(2)}\right),
\end{equation}
where $\gamma_q^{(1)}=\left(\cos q_x+\cos q_y\right)/2$ and $\gamma_q^{(2)}=\cos q_x\cos q_y$. $C_{n,m}$ 
denotes the correlation functions. Assuming 
rotational symmetry, i.e., $\langle S_i^z\rangle=0$, they read 
$C_{n,m}=C_{\bf R}=\langle S_{\bf R}^-S_0^+\rangle=2\langle S_{\bf R}^zS_0^z\rangle$ with ${\bf R}=n{\bf e}_x+m{\bf e}_y$. 
Calculating the second derivative $-\ddot{S}_q^+$, an approximation as indicated in 
Refs.~\onlinecite{haertel08,haertel10,kondo,shimahara91,SSI94,barabanov94,ihle2001,winterfeldt97,schindelin1999,Yu2000,canals2002,schmal2004,schmal06,junger3d,ihle2008,barabanov2009,barabanov2011} is used which implies 
the decoupling scheme
\begin{equation} \label{decoup}
S_i^+S_j^+S_k^-=\alpha_{i,k}\langle S_i^+S_k^-\rangle S_j^++\alpha_{j,k}\langle S_j^+S_k^-\rangle
S_i^+,
\end{equation} 
where the quantities
$\alpha_{i,k}$ are vertex parameters introduced to improve the
decoupling scheme.
In the vicinity of  $J_2^c$ a spin nematic phase could be
present, and the decoupling scheme (\ref{decoup}) might be not appropriate.
Therefore, we restrict our consideration to sufficiently large values of
$J_2$, $J_2>0.7|J_1|$, where the
semi-classical antiferromagnetic collinear stripe
GS LRO is present.
Since  an $\alpha_{i,k}$ is a function of the lattice vector ${\bf R}_i-{\bf R}_k$ 
connecting the sites $i$
and $k$, in what follows we use the same notation as for the correlation
functions $C_{n,m}$, i.e. the vertex parameter $\alpha_{n,m}$ belongs to the
lattice vector ${\bf R}=n{\bf e}_x+m{\bf e}_y$.  
We obtain $-\ddot{S}_{\bf q}^+=\omega_{\bf q}^2S_{\bf q}^+$ and
\begin{equation}\label{Greenfunktion}
  \chi_{\bf q}^{+-}\left(\omega\right)=
-\langle\langle S_{\bf q}^+;S_{-{\bf q}}^-\rangle\rangle_\omega=\frac{M_{\bf q}}{\omega_{\bf q}^2-\omega^2},
\end{equation}
with
\begin{align}\label{omegaq}
  \omega_{\bf q}^2&= 2\sum_{k,l(=1,2)}J_kJ_l\left(1-\gamma_{\bf q}^{(k)}\right)\times\nonumber\\
  &\left[K_{k,l}+8\alpha_{1,k-1} C_{1,k-1}\left(1-\gamma_{\bf q}^{(l)}\right)\right] ,
\end{align}
where $K_{1,1}=1+2\left(2\alpha_{1,1}C_{1,1}+\alpha_{2,0}C_{2,0}-5\alpha_{1,0}C_{1,0}\right)$, 
$K_{2,2}=1+2\left(2\alpha_{2,0}C_{2,0}+\alpha_{2,2}C_{2,2}-5\alpha_{1,1}C_{1,1}\right)$, 
$K_{1,2}=4\left(\alpha_{1,2}C_{1,2}-\alpha_{1,0}C_{1,0}\right)$, and $K_{2,1}=4\left(\alpha_{1,0}C_{1,0}+\alpha_{1,2}C_{1,2}-2\alpha_{1,1}C_{1,1}\right)$. 
The correlation functions $C_{n,m}$ are calculated using
the spectral theorem,\cite{elk} 
\begin{equation}\label{GF-Cq}
  C_{\bf q}=\langle S_{\bf q}^+S_{\bf q}^-\rangle=\frac{M_{\bf q}}{2\omega_{\bf q}}\left[1+2n\left(\omega_{\bf q}\right)\right],
\end{equation}
where $n(\omega_{\bf q})=\left(e^{\omega_{\bf q}/T}-1\right)^{-1}$ is the Bose function.\\
Magnetic LRO is reflected by a non-vanishing condensation
term $C$ (see, e.g., Refs.~\onlinecite{shimahara91,winterfeldt97,ihle2001}) according to 
$C_{\bf{R}}=\frac{1}{N}\sum_{{\bf q}\neq{\bf Q}}C_{\bf q}e^{i{\bf q}\bf{R}}+Ce^{i{\bf Q}\bf{R}}$ 
where ${\bf Q}$ denotes the magnetic ordering vector. The magnetic order
parameter (sublattice magnetization) is calculated as
\begin{equation}
  m^2=\frac{3}{2}\frac{1}{N}\sum_{\bf R}C_{\bf R}e^{-i{\bf Q}{\bf R}}=\frac{3}{2}C.
\end{equation}
The corresponding static susceptibility is given by 
$\chi_{\bf Q}=
\frac{1}{2}\lim_{{\bf q}\to {\bf Q}}\chi_{\bf q}^{+-}\left(\omega=0\right)$. 
The uniform static susceptibility is 
$\chi_{0}=\frac{1}{2}\chi^{+-}_{\bf q=0}(\omega=0)$.
The correlation length is obtained by an expansion of the static susceptibility around 
the magnetic ordering vector ${\bf Q}$, $\chi^{+-}_{{\bf Q}+{\bf q}}\approx
\chi^{+-}_{{\bf
Q}}\left(1-\xi_x^2q_x^2-\xi_y^2q_y^2\right)$, see, e.g.,
Refs.~\onlinecite{winterfeldt97,ihle2001,schmal2004,haertel08,haertel10}. 
From the on-site correlator $\langle S_i^-S_i^+\rangle$ and the operator identity $S_i^-S_i^+=\frac{1}{2}+S_i^z$ 
we get the sum rule 
\begin{equation}\label{sumrule}
C_0=\frac{1}{N}\sum_{\bf q}C_{\bf q}=\frac{1}{2}.
\end{equation}
Considering the collinear stripe phase we have two equivalent magnetic ordering
vectors, ${\bf Q}_{1}=\left(0,\pi\right)$ and ${\bf Q}_{2}=\left(\pi,0\right)$. To preserve
square-lattice  symmetry we
follow Ref.~\onlinecite{ihle2001} and calculate the correlation functions as
$C_{n,m}=\left(C^{(1)}_{n,m}+C^{(2)}_{n,m}\right)/2$ with $C^{(i)}_{n,m}=\frac{1}{N}\sum_{{\bf q}\neq{\bf Q}_{i}}C_{\bf q}e^{i{\bf q}\bf{R}}
+Ce^{i{\bf Q}_{i}\bf{R}}$. 
Note that $C$ is the same for ${\bf Q}_1$ and ${\bf Q}_2$. 
Analogously, we consider  $\chi_{{\bf Q}+{\bf q}}=\left(\chi_{{\bf Q}_1+{\bf q}}+\chi_{{\bf Q}_2+{\bf q}}\right)/2$
to get the  
correlation length by expansion of $\chi_{{\bf Q}+{\bf q}}$ for small ${\bf
q}$ yielding 
\begin{align}\label{corrlength}
  \xi^2&=\frac{2J_2C_{1,1}}{4J_1C_{1,0}+8J_2C_{1,1}}+\frac{A}{\Delta},
\end{align}
where 
$\Delta=2J_1^2K_{1,1}+4J_2^2K_{2,2}+2J_1J_2\left(K_{1,2}+2K_{2,1}\right)+16\left(J_1+2J_2\right)
\left(J_1\alpha_{1,0}C_{1,0}+2J_2\alpha_{1,1}C_{1,1}\right))$
and
$A=-J_2^2\left(32\alpha_{1,1}C_{1,1}+K_{2,2}\right)-4J_1J_2\left(3\alpha_{1,0}C_{1,0}+\alpha_{1,2}C_{1,2}\right)$.

In the GS, LRO may exist, that is, $C \ne 0$.   
 $C$ is determined by\cite{shimahara91,schindelin1999,schmal06}
$\chi^{-1}_{{\bf Q}_{1}}=\chi^{-1}_{{\bf Q}_{2}}=0$ with
\begin{equation} \label{condens}
  \chi_{{\bf Q}_i}=-\frac{-4J_1C_{1,0}-8J_2C_{1,1}}{\Delta} .
\end{equation}

\begin{figure}[ht]
  \begin{center}
    \includegraphics[height=6.5cm]{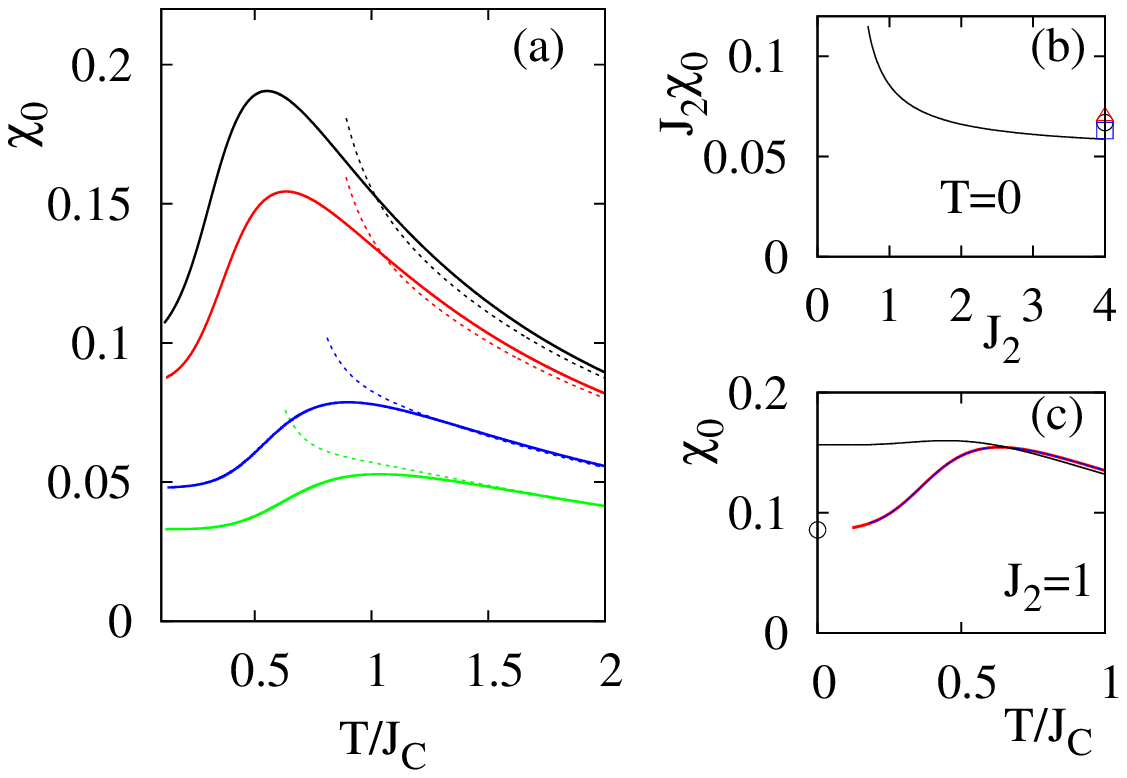} 
  \caption{Uniform susceptibility $\chi_0$ of the $J_1$-$J_2$ model for
 ferromagnetic $J_1=-1$.\\
(a) Temperature dependence of the uniform susceptibility $\chi_0$ for
$J_2=0.9$ (black), $J_2=1.0$ (red), $J_2=1.5$ (blue), and $J_2=2.0$
  (green) calculated by RGM (solid lines). The temperature $T$ is scaled by
  $J_{C}= \sqrt{J_1^2 + J_2^2}$. 
  For comparison, eighth-order high-temperature expansion results
  (Refs.~\onlinecite{Rosner_HTE,schmidt11}) are shown (dashed lines).\\
  (b) Ground-state
  values  for $J_2 \chi_0$ vs. $J_2$. 
  The susceptibility data for the square-lattice antiferromagnet
  (quantum Monte Carlo\cite{runge92} - black circle, 
  CCM\cite{farnell09} - red triangle, third-order spin-wave theory\cite{hamer92}
  - blue square)
  are also shown as data points at $J_2=4$.\\
  (c)  Comparison of the temperature dependence of the uniform susceptibility $\chi_0$
  for $J_2=1$ calculated using three different choices of the vertex parameters (red line - two vertex parameters with 
  the ansatz
  (\ref{var2})  and $\gamma=0.005$, blue line - two vertex parameters with 
  fixed ratio
  $\frac{\alpha_2}{\alpha_1}(T)=\frac{\alpha_2}{\alpha_1}(0)$ (i.e.  ansatz
  (\ref{var2}) with $\gamma=0$), black line -
  only one vertex parameter). 
  Note that the red and the blue lines practically coincide. Note further
  that the red line extends down to slightly lower temperatures.
  The black circle
  at $T=0$ shows the zero-temperature RGM value for $\chi_0$ calculated
  with two vertex parameters.}
 \label{fig1}
  \end{center}
\end{figure}

Next we have to discuss the choice of the vertex parameters
$\alpha_{n,m}$. Obviously, there are five different $\alpha_{n,m}$ in
Eq.~(\ref{omegaq}) which have to be determined together with the 
corresponding correlation functions $C_{n,m}$. In addition, at zero
temperature the condensation term $C$ (describing magnetic LRO) has to be
considered.   
To determine these quantities  we can use  the Fourier transformation of
Eq.~(\ref{GF-Cq}) providing five equations for $C_{n,m}$.
Moreover, at zero temperature we use $\chi^{-1}_{{\bf
Q}_{1}}=\chi^{-1}_{{\bf Q}_{2}}=0$ to calculate $C$, see above.
Finally, only one equation, namely the sum rule (\ref{sumrule}), is left to find the vertex
parameters. Hence, we have to introduce further approximations.
In the case of a ferromagnetic GS ($J_2<J_2^c$) all correlation functions behave
quite similar and a reasonable approximation is to set $\alpha_{n,m}=\alpha$, see,
e.g., Refs.~\onlinecite{shimahara91,haertel08} and \onlinecite{haertel10}.
This simple approximation was also used in Ref.~\onlinecite{kondo} applying the RGM to antiferromagnets.
However, in the antiferromagnetic regime the correlation functions carry
different signs, and setting $\alpha_{n,m}=\alpha$ leads to poor results at
low temperatures.
A significant improvement of the RGM results for antiferromagnets can be achieved by introducing     
two independent vertex parameters.\cite{shimahara91,ihle2001,schmal06} 
This requires, however, an additional external input to get one more equation.
To take into account the dominant character of $J_2$
in the collinear stripe phase, we set $\alpha_{1,1}=\alpha_1$  and
$\alpha_{n,m}=\alpha_2$, $(n,m) \ne
(1,1)$.
Since the low-temperature properties of the model  are related to
excitations
above the GS, a realistic description of the GS is
necessary. Therefore, in the present paper
we use, as an additional external input, the GS sublattice magnetization
calculated by the CCM.\cite{richter10} 
Thus, describing GS magnetic ordering properly, we may expect that the
RGM provides also a reasonable description of the 
low-temperature properties of the model. 
This input yields the required additional equation to determine the two
independent vertex parameters $\alpha_1$ and $\alpha_2$ at $T=0$.
As a result we can also calculate the uniform static susceptibility
$\chi_{0}$ at $T=0$, cf. Fig.~\ref{fig1}. 

For finite temperatures we need a reasonable ansatz for the temperature
dependence of the ratio $\alpha_2/\alpha_1$ of the vertex parameters, see, e.g.  
Refs. \onlinecite{winterfeldt97,shimahara91,ihle2001} and
\onlinecite{junger3d}.
We have tested several ansatzes to a get a proper description of
thermodynamic quantities in the whole temperature range, see below.
To solve the system of RGM equations we use Broyden's method,\cite{NR3} which yields the
solutions with a
relative error of about $10^{-8}$ on the average. The momentum integrals are
done by Gaussian integration.
To find the numerical solution of the equations for $T> 0$,  we start
at high temperatures and decrease $T$ in small steps. Below a certain
(low) temperature $T_0(J_2)$ no solutions of the RGM equations (except at
$T=0$) could be found,  since the
quantity $\Delta(T,J_2)$ in Eqs.~(\ref{corrlength}) and (\ref{condens})
becomes exponentially small
which leads to numerical instabilities.
As expected, 
at large temperatures the concrete choice of the
ratio $\frac{\alpha_2}{\alpha_1}(T)$ becomes irrelevant, and even the simple approximation 
$\alpha_{n,m}=\alpha$ yields results for $\chi_0(T)$ which coincide with the data from the high-temperature
expansion. 
At low temperatures a resonable ansatz for the
ratio $\alpha_2/\alpha_1$  should (i) provide numerical data down to
sufficiently low temperatures and (ii) yield coincidence of
$\chi_0(T=0)$ determined by using the CCM input and $\lim_{T\to 0}\chi_0(T)$
calculated with the ansatz for $\frac{\alpha_2}{\alpha_1}(T)$. 
The simplest way is to fix the ratio $\frac{\alpha_2}{\alpha_1}$ to its
value at $T=0$.
Tracing the RGM solution to very low temperatures we find that the ansatz
\begin{equation}\label{var2}
\frac{\alpha_2(T)}{\alpha_1(T)}=1+\left(\frac{\alpha_2(0)}{\alpha_1(0)}-1\right)e^{-\gamma
T} \; , \;\gamma \ge 0
\end{equation}   
with the tiny exponent $\gamma=0.005$ is more appropriate to get numerically
stable solutions at low $T$. Note,
however, that our results are not noticeably influcenced by the choice of
$\gamma$,  see also Fig.~\ref{fig1}.
In what follows we use this ansatz to solve the RGM equations with two vertex
parameters.

\section{Results and Discussion}\label{secIII}
In what follows we fix the ferromagnetic NN exchange to $J_1=-1$. We focus
on sufficiently large values of $J_2$, $J_2>0.7$, where a possible nematic GS
phase is not present, and, rather the GS exhibits semi-classical collinear magnetic
LRO.
Moreover, the  experimental data for layered oxovanadates\cite{kaul04,nath08,carretta2011} correspond to this parameter regime.
We follow Ref.~\onlinecite{shannon04} and
use as a characteristic energy scale $J_C=\sqrt{J_1^2+J_2^2}$, see also
Refs.~\onlinecite{nath08,carretta2011}.


In Fig.~\ref{fig1}(a) the temperature dependence of the uniform susceptibility $\chi_0$ is
shown.
For comparison we also present the results of  the eighth order high-temperature
expansion,\cite{schmidt11} which agree with the RGM data at large $T$.
In  Fig.~\ref{fig1}(b)  the GS results for the susceptibility
$\chi_0(T=0)$ are presented. 
Since for large $J_2$ the $J_1$-$J_2$ model corresponds to a system of two
inter-penetrating square-lattice
antiferromagnets with coupling strength $J_2$, our RGM data for $\chi_0(T=0)$ 
can be compared with available GS results for $\chi_0(T=0)$ of the square-lattice
antiferromagnet,\cite{runge92,hamer92,farnell09} see data points at $J_2=4$
in Fig.~\ref{fig1}(b).  
For large $J_2$, the dependence of $J_2\chi_0(T=0)$ on $J_2$ is weak down to $J_2 \sim 1$. 
A noticeable upturn of $J_2\chi_0(T=0)$ for small $J_2$ may indicate the
approach to the ferromagnetic phase.  
In Fig.~\ref{fig1}(c) we compare different
choices of the vertex parameters to solve the RGM equations, see the
discussion in Sec.~\ref{sec2}. 
Obviously, the use of only one vertex parameter by setting $\alpha_{n,m}=\alpha$ 
leads to poor results for $T \lesssim 0.6 J_{C}$. 
On the other hand, the two choices of the parameter $\gamma$
in the ansatz (\ref{var2}) lead to almost identical $\chi_0(T)$ curves.

\begin{figure}[ht]
  \begin{center}
    \includegraphics[height=10cm]{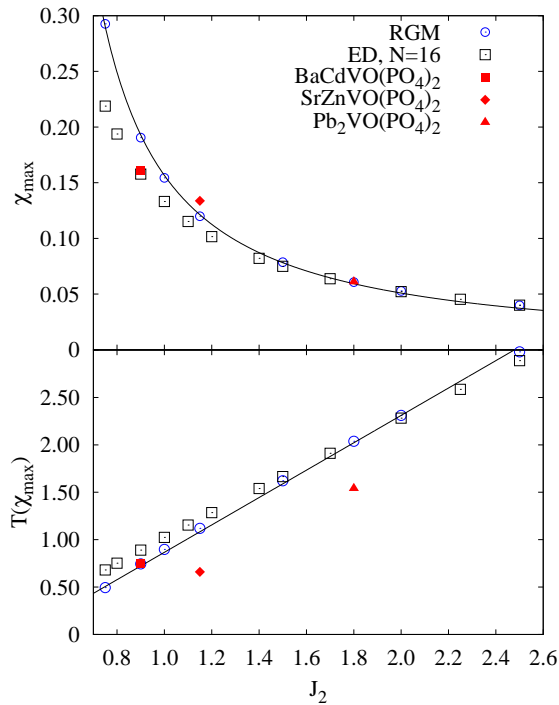}
  \caption{Height and position of the maximum of the susceptibility
$\chi_0(T)$ for the $J_1$-$J_2$ model with $J_1=-1$ in dependence
on the frustration parameter $J_2$. The blue circles are the RGM results,
the open squares show the ED data for $N=16$, the
lines correspond to Eqs.~(\ref{fit1}) and (\ref{fit2}), and the colored
filled
symbols correspond to  BaCdVO(PO$_4$)$_2$ ($J_2/|J_1|\approx 0.9$)\cite{nath08},
SrZnVO(PO$_4$)$_2$ ($J_2/|J_1|\approx 1.15$)\cite{carretta2011}, 
 and Pb$_2$VO(PO$_4$)$_2$
($J_2/|J_1|\approx 1.8$).\cite{kaul04} 
}\label{fig2}
  \end{center}
\end{figure}

For the comparison with experimental data on oxovanadate
compounds, such as BaCdVO(PO$_4$)$_2$,
SrZnVO(PO$_4$)$_2$,
 and Pb$_2$VO(PO$_4$)$_2$
\cite{kaul04,nath08,carretta2011},  the height $\chi_{max}$ and  the
position $T(\chi_{max})$ of the maximum in the $\chi_0(T)$ curve 
are interesting features.
However, in these compounds, due to a weak interlayer coupling, 
a phase transition to magnetic long-range order at a critical temperature 
$T_N$ was detected,\cite{kaul04,nath08,carretta2011} 
where $T_N/|J_1| \sim  0.28, 0.35$, and $0.48$ was found  for
BaCdVO(PO$_4$)$_2$,
SrZnVO(PO$_4$)$_2$,
 and Pb$_2$VO(PO$_4$)$_2$, respectively.
Since in our paper we deal  with a strictly two-dimensional model, such a
 comparison is reasonable only in the paramagnetic phase of the compounds,
 i.e. at $T>T_N$, where
two-dimensional spin physics dominates the thermodynamic behavior.
Indeed, the relevant temperatures $T(\chi_{max})$ are well above $T_N$ for all three
oxovanadates, namely $T(\chi_{max})/|J_1| = 0.75, 0.66$,
and $1.54$ for BaCdVO(PO$_4$)$_2$,
SrZnVO(PO$_4$)$_2$,
 and Pb$_2$VO(PO$_4$)$_2$, respectively, see Fig.~\ref{fig2}.

We mention, that $\chi_{max}$ and $T(\chi_{max})$ have been presented as
fuctions of $\tan^{-1}(J_2/J_1)$ for the whole parameter space of $J_1$ and
$J_2$ in
Ref.~\onlinecite{shannon04} using full ED for $N=8,16$ and $20$
sites, see Fig.~17 in
Ref.~\onlinecite{shannon04}. 
We have recalculated the corresponding ED data for $N=16$.  In Fig.~\ref{fig2}
we show the ED and the RGM results for $\chi_{max}$ and $T(\chi_{max})$  
as functions of $J_2$ in the parameter region  $J_2>0.7$ considered in our
paper.

The RGM data points are well described by the relations
\begin{eqnarray}
      \chi_{max}&=a \left(J_2-J_2^c\right)^{-\nu}, \label{fit1} \\
   T(\chi_{max})&=b(J_2-J_2^c) \label{fit2} 
\end{eqnarray}
with $a= 0.0872 $, $\nu = 1.146$, $b=1.444$, and  $J_2^c=0.4$.
Note, however, that the divergence of $\chi_{max}$ at
$J_2^c=0.4$ suggested by Eq.~(\ref{fit1}) might be absent in the considered model, since
our approach is not designed for $J_2 \sim J_2^c$.
Using experimental data of the susceptibility for quasi-two-dimensional
frustrated square-lattice magnets  
as well as the reported exchange constants $J_1$ and $J_2$ we can compare  our theoretical data
directly with experiment, see Fig.~\ref{fig2}. Obviously, theory and
experiment agree well, particularly for $\chi_{max}$.
Hence, our equations (\ref{fit1}) and (\ref{fit2}) can be used 
to get information on the ratio $J_2/|J_1|$ from susceptibility measurements.

\begin{figure}[ht]
  \begin{center}
    \includegraphics[height=6cm]{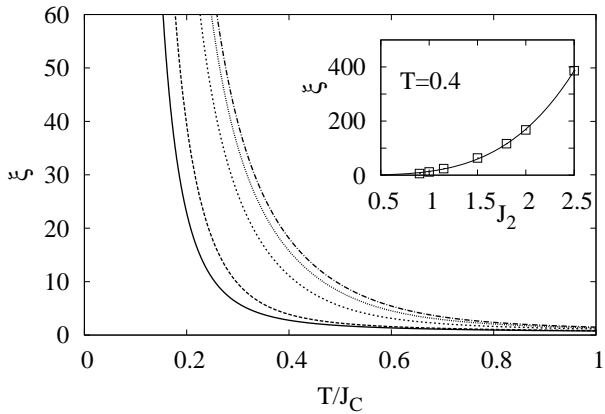}
  \caption{Temperature dependence of the correlation length $\xi$
 (in units  of the lattice spacing) for different values of
the frustration parameter $J_2$=0.9, 1.0, 1.5, 2.0 and 2.5
(from left to right).
The temperature $T$ is scaled by
  $J_{C}= \sqrt{J_1^2 + J_2^2}$.
The inset shows the correlation length for $T=0.4$ in dependence on the
frustration parameter $J_2$. The solid line represents a fit
to the data points (open squares).
}\label{fig3}
  \end{center}
\end{figure}

\begin{figure}[ht]
  \begin{center}
    \includegraphics[height=7cm]{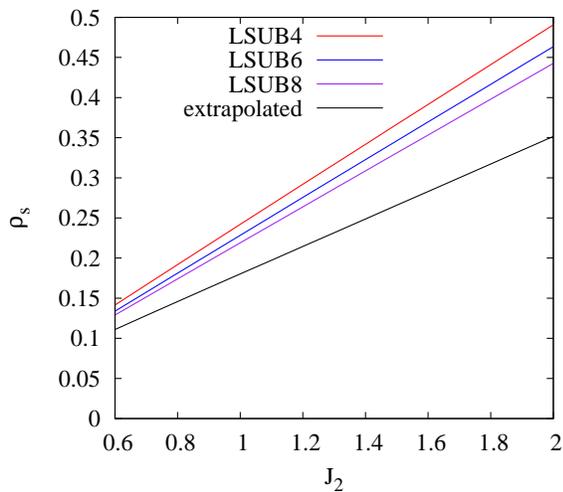}
  \caption{Spin stiffness $\rho_s$
  of the $J_1$-$J_2$ model for
 ferromagnetic $J_1=-1$ as a function of $J_2$ calculated by the CCM.
}\label{fig4}
  \end{center}
\end{figure}
\begin{figure}[ht]
  \begin{center}
    \includegraphics[height=7cm]{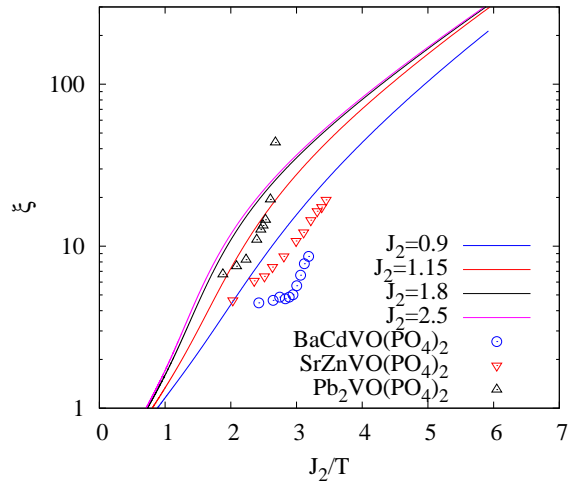}
  \caption{Correlation length $\xi$ (logarithmic scale)  in
 units  of the lattice spacing vs. $J_2/T$ calculated using RGM (lines)
for various values of the frustration parameter $J_2$.  For comparison we
also show experimental data\cite{carretta2011} for BaCdVO(PO$_4$)$_2$
($J_2/|J_1|\approx 0.9$),
SrZnVO(PO$_4$)$_2$ ($J_2/|J_1|\approx 1.15$), and Pb$_2$VO(PO$_4$)$_2$ ($J_2/|J_1|\approx 1.8$).
}\label{fig5}
  \end{center}
\end{figure}

Next we discuss  the correlation length $\xi$.
Its temperature dependence  is depicted in Fig. \ref{fig3} for different values of
the  frustration parameter $J_2$.
With increasing NNN exchange $J_2$ the rapid increase in $\xi$ is
shifted to larger temperatures. As shown in the inset of
Fig.~\ref{fig3}, at a certain fixed temperature, $\xi$ decreases rapidly  with
decreasing $J_2$.

The exponential low-temperature divergence of
$\xi \propto e^{A/T}$ for two-dimensional Heisenberg
magnets with NN interactions is determined by the spin stiffness $\rho_s$, i.e. $A \propto \rho_s$,
see e.g.
Refs.~\onlinecite{kopietz,auerbach,chakravarty,hasenfratz,greven,haertel10}.
As it has been recently reported,\cite{haertel10}
for small $J_2$, where a ferromagnetic GS is present,
the relation $\xi \propto e^{a\rho_s/T}$ also holds if the NNN exchange $J_2$ is
included. In this case the stiffness was 
obtained from the RGM dispersion relation\cite{haertel10}
as $\rho_s^{FM}=-(J_1+2J_2)/4$.
For the antiferromagnetic collinear stripe GS  phase present at large $J_2$
one may expect that the stiffness also determines the exponential
divergence at small $T$. However,
the determination of $\rho_s$ is more difficult.
Here we use the CCM\cite{bishop04,krueger06,darradi08} to
provide data for $\rho_s$.
To calculate $\rho_s$ within the CCM we follow strictly
Refs.~\onlinecite{krueger06} and \onlinecite{darradi08} and do not explain
details of the calculation.
The stiffness as a function of $J_2$ is shown in Fig.~\ref{fig4} for
various levels of CCM approximations, LSUB$n$, as well as 
extrapolated data.\cite{lsubn} 
The obvious (almost) linear $J_2$-dependence of $\rho_s$ is well
described for the extrapolated CCM data by  $\rho_s \approx 0.175 J_2$.
Hence, it seems to be reasonable to show the temperature dependence of the correlation
length,
in addition to Fig.~\ref{fig3}, as a function $\ln \xi(J_2/T)$, see
Fig.~\ref{fig5}.
First we notice that the experimental data reported in
Ref.~\onlinecite{carretta2011} agree reasonably well with our RGM results.
Secondly, it is obvious that for large values of $J_2 \gtrsim 1.5$  the  $\ln \xi(J_2/T)$ curves
almost coincide. The small deviations can be attributed to a temperature
dependent prefactor in front  of the exponential
term.\cite{kopietz,auerbach,chakravarty,hasenfratz,greven,haertel10}
However, for $J_2=1.15$ and $J_2=0.9$ the theoretical as well as the
experimental data 
show deviations from the behavior suggested by the stiffness
data.

\section{Summary}\label{secIV}
Using second-order Greens function technique we have calculated the uniform susceptibility
$\chi_0$ and the correlation length $\xi$ of 
the frustrated $J_1$-$J_2$ square-lattice Heisenberg ferromagnet in the collinear
antiferromagnetic regime present for large values of $J_2/|J_1|$.   
We have derived simple power laws for the height and the position of the 
maximum in the $\chi_0(T)$ curve as functions of $J_2$.
We have found that our theoretical data agree reasonably well with recent experiments on
oxovanadates.  

\vspace*{1cm}

{\bf Acknowledgment}
The authors are indebted to O. Derzhko for critical reading of the
manuscript.\\
J.R. and S.-L.D. thank the
DFG for financial support (grants
DR269/3-3  and RI615/16-3).


\begin{thebibliography}{99}
\bibitem{lsubn} 
The LSUB$n$ approximation scheme is widely used in CCM calculations, see
e.g. Refs.\onlinecite{bishop04,krueger06,darradi08}.  
Within the LSUB$n$ scheme
all multispin correlations
over all distinct locales on the lattice defined by $n$ or fewer contiguous
sites are taken into account.
Since the LSUB$n$ approximation becomes exact for $n \to \infty$,
it is useful to extrapolate the ''raw'' LSUB$n$ data to $n \to \infty$ by
$\rho_s(n)=c_0+c_1 (1/n) + c_2(1/n)^2$.
\end{thebibliography}
\end{document}